# Using rule engine in self-healing systems and MAPE model.

Zahra Yazdanparast

*School of Electrical and Computer Engineering,*
*Tarbiat Modares University,*
*Tehran, Iran*
zahra.yazdanparast@modares.ac.ir

**Abstract**

Software malfunction presents a significant hurdle within the computing domain, carrying substantial risks for systems, enterprises, and users universally. To produce software with high reliability and quality, effective debugging is essential. Program debugging is an activity to reduce software maintenance costs. In this study, a failure repair method that uses a rule engine is presented. The simulation on mRUBIS showed that the proposed method could be efficient in the operational environment. Through a thorough grasp of software failure and the adoption of efficient mitigation strategies, stakeholders can bolster the dependability, security, and adaptability of software systems. This, in turn, reduces the repercussions of failures and cultivates increased confidence in digital technologies.

## 1. Introduction

Dynamically adaptive systems must be able to adapt its structure and/or behavior in response to changes. Such a system must be reliable and always available to remain useful. During execution, two types of changes may occur: 1) the operating environment of the software changes and 2) new requirements that were not predicted at design time appear at execution time. For this purpose, dynamic reconfiguration should be applied at runtime, whenever necessary. Dynamic reconfiguration is the conversion of the current system configuration to another at runtime. Reconfiguration can be anticipated or unanticipated. Predictive reconfiguration is caused by expected changes in requirements, so it can be programmed ahead of time. While unanticipated configuration is caused by unexpected requirements change, therefore, it is not possible to predict or plan reconfiguration before the change occurs [1].

Self-adaptive systems demonstrate an impressive capacity to independently adjust their operations in reaction to shifts in their surroundings or internal fluctuations [2]. These adjustments are capable of taking place during runtime, enabling the system to enhance its efficiency, utilization of resources, and ability to withstand challenges [3]. Self-healing systems strive to deliver resilience and fault tolerance by autonomously identifying, diagnosing, and rectifying failures [4, 5]. Such systems prove especially vital in critical and ever-changing environments where unforeseen faults may arise. Nevertheless, assessing their efficacy continues to present a hurdle [6].

The MAPE (Monitoring, Analysis, Planning, Execution) control loop model plays a pivotal role in enabling self-adaptation within software systems [7]. Drawing inspiration from the autonomic nervous system, this model has been recognized as a fundamental element in overseeing and directing autonomous and self-adjusting systems [8]. In this study, we focus on Planning and provide a solution for choosing the appropriate healing for the system.

## 2. Related work

In [9] introduced a solution for testing and self-healing of cloud resources. The proposed solution is based on a multi-agent architecture that uses the elasticsearch, logstash and kibana(ELK) and a rule-based self-healing algorithm implemented with the Drools rule engine and the spring boot framework. Preliminary tests have shown that the proposed solution meets its objectives in terms of maintaining service level agreements, reducing setup costs, high performance, and operability. The main idea of the rule-based self-healing algorithm is implemented using Drools. Drools works in two circles of exploration (training) and circle of productivity. In the exploration phase, the output of the execution rejection and the expert's knowledge are used to create rules. The productivity loop is automatically activated after the exploration phase is completed (reaching the defined threshold) and if it encounters an issue during the self-healing process that it cannot handle, it collects the appropriate rejections and sends a notification to the relevant engineer to take care of that failure. The solution provided only solves known problems and if a new problem arises, it is referred to a human expert.

The aim of [10] is to provide a self-healing method based on replication that can perform data synchronization on different nodes in high failure rate environments. Each node has a set of agents. Each agent collects data from each node and coordinates them. Then it moves between different nodes and copies the data on them. Finally, the data of all the considered nodes will be the same. In case of the failure of an agent, the node creates a new agent. According to the end of the time limit, each node notices the failure of the agent and creates a new agent. There are two problems here: how much timeout should be and if the system is large, the number of messages will be large. The proposed method is that the agents move between different nodes to copy the data on them. When an agent leaves a node, it sends a departing message and the destination node to the source node, and when it reaches the destination node, it sends a freeresp message to the source node. The time interval between receiving these two messages is considered as a time limit. If the node does not receive the freeresp message after the timeout, it creates a new agent. If a freeresp message is received after creating a new agent, the new agent is not deleted, but the next incoming agent is deleted after copying its information. The result of the article has been evaluated using four complex network models: small world, scale-free, hub, and community network.

The [11] presents a framework for the self-healing of cloud services. The technique of this paper is divided into two stages: the first stage is the usefulness and reliability of the service, in which the probability of service failure is calculated. The second step is program and run-time evaluation, where a recovery program is built to run when a failure occurs. When a runtime failure occurs, the system first uses WS-BPEL, a set of exception-handling strategies, to repair the failure. If the failure is repaired, the service continues to run, otherwise, the system goes to the scheduling module, which generates and executes the optimal schedule. The planning module consists of three parts: the decision component, the confirmation component, and the planning component.

The [12] proposes a framework for self-healing of cyber-physical systems. The self-healing service implemented by failure detection (collection and analysis) and improvement (planning and action) of the components. Structural adaptation changes the structure of the system at runtime, removing, adding, and rearranging components. In contrast, a parameter adaptation changes the behavior of a component by adapting its parameters. ORR focuses on structural adaptation in real-time service-based systems, and when a service fails, it can design a replacement for it. An alternative algorithm uses a knowledge base that contains relationships between features in the CSP, called an ontology, and defines additional CPS runtime information. While the ontology contains only static information, dynamic changes such as adding, removing, and adapting services at runtime are modeled by a table called Service-to-Ontology Mapping (SOM).

## 3. Suggested approach

Different types of failures are defined by the designer so that the type of failure can be recognized in the analysis stage. Types of failures at the architectural level include the following [13, 14]:

- The component is failed and enters the unknown state (CF1).
- The number of component exception type failure that are greater than the specified threshold (CF2).

- The component is removed from the running software architecture (CF3).
- The connection between two components of the running software architecture has been removed (CF4).

To repair the failure, rule-based design is performed in this study. It is necessary to specify these rules in advance by the designer and put them in the rule engine. These rules, called adaptive strategies (AS), include the following [13, 14]:

- Restart component (AS1)
- Redeployment of the component (AS2)
- Reestablishing the relationship between two components (AS3)
- Replacing the component with a new sample of the same type (AS4)

### 3.1. Checking the root of the failure

In this section, the dependent components are first examined. The reason for checking the dependent components is that the root of the declared failure may not be the failed component, and the component that is declared failed does not have a problem. However, due to a failure in the components it uses, this component has been declared damaged. In a way, the results of the previous components cause the failure of the component that is declared as a failed component. A counter is utilized to identify failures in the dependent components. In this way, every time a failed component is declared, the counter of each of its dependent components increases. If the counter value of any component exceeds a certain threshold, it can be concluded that the component is damaged and it will cause a failure to be declared in its child components. Therefore, it is necessary to declare this failed component and determine the type of failure and design for it.

### 3.2. Plan

After detecting the type of failure and determining the failed component, in the proposed method, the repair is carried out according to Figure 1 based on the rule. In the proposed method, the communication between the analysis section and the design section is performed using socket programming. The failure of the component is sent from the analysis to the plan to choose the appropriate repair. Therefore, analysis and plan can be implemented in a distributed manner. Then, the rule engine is used to select the appropriate repair plan among different repair plans according to the type of failure. The rule engine separates program code and rules, so adding a new rule is easy.

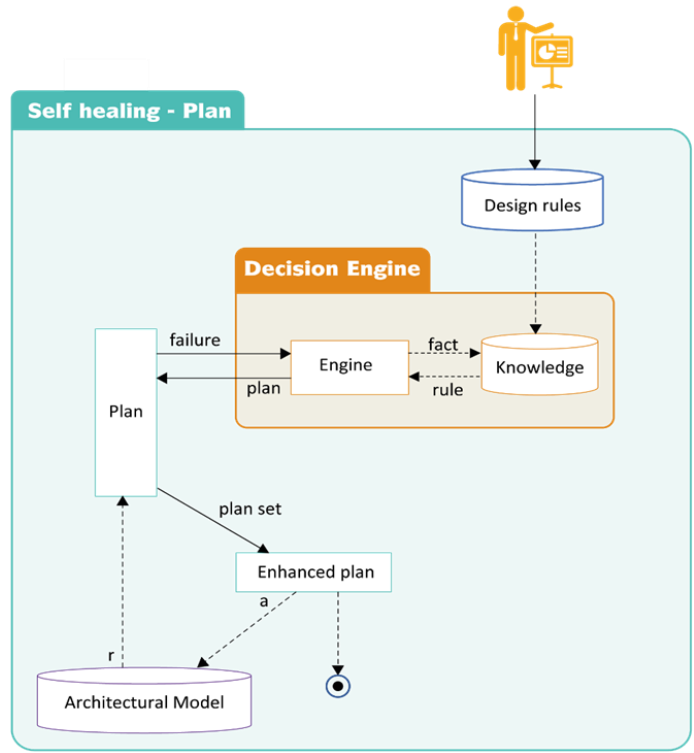

Figure 1: Plan.

## 4- Evaluation of the results

In the plan, appropriate repair of the failure is selected. Drools rule engine is applied in this section. Socket programming is used to communicate between analysis and Drools.

After analyzing the failure and determining its type, using the Drools tool, the appropriate plan is selected for repair. Finally, the plan is executed to restore the system. After completing the self-healing cycle, the mRUBiS simulator verifies the architecture to ensure that there are no further failures. Then, the system continues to work and after a certain time (between 100 and 500 milliseconds) the second crash is injected. The self-healing loop is executed again and the failure is repaired. The failures are randomly selected from among the existing failures and injected into the architectural model.

### 4.1. Using the rule engine to repair the failure

There are various tools to implement the rule engine, and in this research, the Drools tool was used. In mRubis simulator with CF4 failure, the connection between the two Reputation Service and Query Service components is interrupted. After identifying the failure in the analysis, it is necessary to select the appropriate plan for this failure. Drools is implemented in a distributed manner. Therefore, by programming the socket, it is connected to Drools and according to the detected failure, the appropriate program is selected for repair.

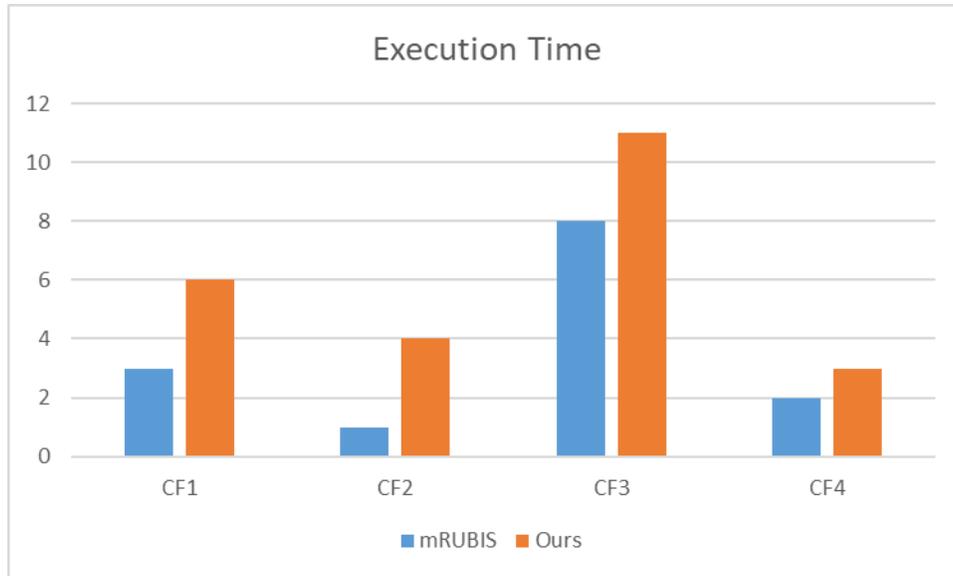

Figure 2: The result of the CF4 repair.

### 4.2. Declaring the dependent component as a possible source of failure

In the first scenario, each of the four crash types was executed only once, during which the dependent components were identified for each failed component. However, no analysis was performed in this regard and the self-healing procedure was carried out as before.

In this scenario, the aim is to investigate the effect of dependent components on the failure of a component. That is, if a failed component is declared, is there a possibility that the root cause of the failure is another component? To check this mode, in 20 rounds, different failures are injected into the components. If a component is declared as the root of a failure more than three times (threshold is considered 3), the name of that the component and its impact on failures are stored in a file so that component can be checked. The same component that is declared failed is repaired and the component that is the possible root of the failure is declared for further investigation. In this case, the Query Service component failed three times, and each time the Last Second Sales Item Filter component is stated as the root of the failure.

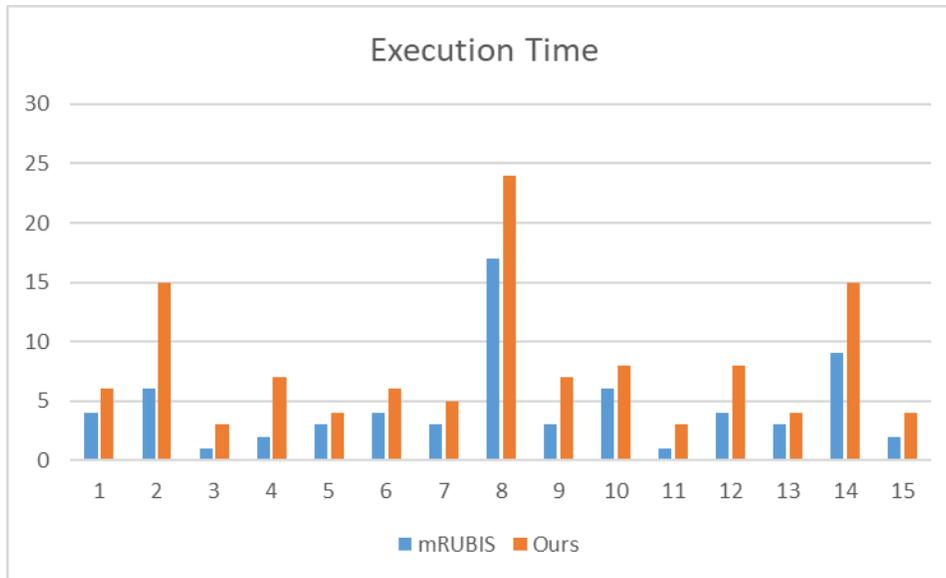

Figure 3: The result of scenario.

## 5. Conclusion

In the realm of computing, software failure is an inevitable reality that organizations must confront. By embracing failure as a natural part of the software lifecycle, organizations can foster a culture of continuous learning and improvement. Through post-mortem analyses, rigorous testing, and proactive measures, software systems can be fortified against future failures. This approach not only minimizes the impact of failures but also promotes greater trust and confidence in digital technologies. In essence, the journey towards software reliability is not about avoiding failure altogether, but rather about embracing it as a catalyst for growth and evolution. In this study, using Drools tool, a distributed approach for planning phase of self-healing software was proposed. the method was evaluated using mRUBIS simulator.